\documentclass[reprint,aps,ams,amsmath,prx,twocolumn,longbibliography,superscriptaddress]{revtex4-1}
\usepackage{graphics}
\usepackage{graphicx}
\usepackage{epstopdf, epsfig}
\usepackage{amsmath}
\usepackage{amssymb}
\usepackage{amsfonts}
\usepackage{bm}
\usepackage{color}
\usepackage{bbm}
\usepackage{hyperref}
\usepackage[normalem]{ulem}
\usepackage{comment}
\usepackage{soul}

\DeclareMathOperator{\sech}{sech}
\newcommand{\be}{\begin{equation}}
\newcommand{\ee}{\end{equation}}
\def\rr#1{(\ref{#1})}

\begin{document}

\title{Photogalvanic effect in hydrodynamic flows of nonreciprocal electron liquids}

\author{E. Kirkinis}
\affiliation{Center for Computation and Theory of Soft Materials, Robert R. McCormick School of Engineering and Applied Science, Northwestern University, Evanston IL 60208 USA}

\author{L. Bonds}
\affiliation{Department of Physics, University of Washington, Seattle, Washington 98195, USA}

\author{A. Levchenko}
\affiliation{Department of Physics, University of Wisconsin-Madison, Madison, Wisconsin 53706, USA}

\author{A. V. Andreev}
\affiliation{Department of Physics, University of Washington, Seattle, Washington 98195, USA} 

\date{September 1, 2025}

\begin{abstract}
We study nonlinear hydrodynamic electron transport driven by an AC electric field. In noncentrosymmetric conductors with broken time-reversal (TR) symmetry the nonlinear flow of such liquids is nonreciprocal, giving rise to a DC current $I^{DC}$ that is quadratic in the amplitude of the AC electric field. This is the hydrodynamic analogue of the linear photogalvanic effect (PGE), which arises in bulk noncentrosymmetric materials with broken TR symmetry. The magnitude of $I^{DC}$ depends on both the properties of the electron fluid and the geometry of the flow, and may be characterized by two dimensionless parameters: the nonreciprocity number $\mathcal{N}$, and the frequency-dependent vibrational number $\mathcal{R}$. 
Due to nonlocality of hydrodynamic transport, at low frequencies of the AC drive, $I^{DC}$ is super-extensive. The AC component of the electric current is likewise strongly affected by nonreciprocity: the hysteretic current-voltage dependence becomes skewed, which can be interpreted in terms of nonreciprocity of the memory retention time.
\end{abstract}

\maketitle

\section{Introduction}

In solid state systems in which the spatial and temporal scales associated with the relaxation of electronic momentum and energy are long compared to those of momentum- and energy-conserving electron-electron scattering, the electron transport can be described using the hydrodynamic approach \cite{Gurzhi1963,Gurzhi1968}. 
Hydrodynamic effects in electron transport have been predicted to occur in strongly-correlated electron systems \cite{Andreev2011}, Dirac liquids in graphene \cite{Muller2009}, coupled electron-phonon systems \cite{Levchenko2020}, and established through advances in the fabrication of high-mobility and low density semiconductor heterostructures, cf. Refs.~\cite{Molenkamp1994,deJong1995,Gao2010}. Signatures of hydrodynamic electron transport have been verified in a number of recent experiments, for comprehensive reviews on the topic see Refs. \cite{LucasFong2018,Narozhny2022,Fritz2024} and references herein.

Flows of electron liquids display properties which have no counterparts in the conventional Newtonian hydrodynamics. For instance, in noncentrosymmetric conductors with broken time-reversal symmetry, the Poiseuille flow of the electron liquid exhibits strongly nonreciprocal (in the sense of Onsager \cite{Onsager1931}) nonlinear two-terminal conductance~\cite{Kirkinis2025}.
For these reasons, in the following we refer to the electron liquid in noncentrosymmetric materials with broken TR invariance as nonreciprocal. 
The enhancement of nonreciprocity of the nonlinear conductance in the hydrodynamic regime arises from the linear velocity-dependent correction to the fluid viscosity \cite{Kirkinis2025}, which is allowed by symmetry in nonreciprocal liquids. 

In this paper we consider nonlinear flows of nonreciprocal electron liquids driven by oscillating electric fields. The AC drive gives rise to a DC current $I^{DC}$, which is the hydrodynamic analogue of the linear photogalvanic effect (DC current induced by linearly-polarized light), arising in noncentrosymmetric media with broken time-reversal symmetry~\cite{Sturman1992,Ivchenko2005,Ivchenko2012}.
In the hydrodynamic regime, the magnitude of the DC current depends not only on the degree of nonreciprocity of the electron liquid but also on the geometry of the flow. We show that the magnitude of the DC current may be expressed in terms of the nonreciprocity number $\mathcal{N}$, which characterizes static nonlinear nonreciprocal flows~\cite{Kirkinis2025}, and the frequency-dependent vibrational number $\mathcal{R}$, which characterizes the extent of vorticity spreading from the sample boundary to the bulk of the flow.

 \begin{figure}[t!]
\includegraphics[height=2.6in,width=4in]{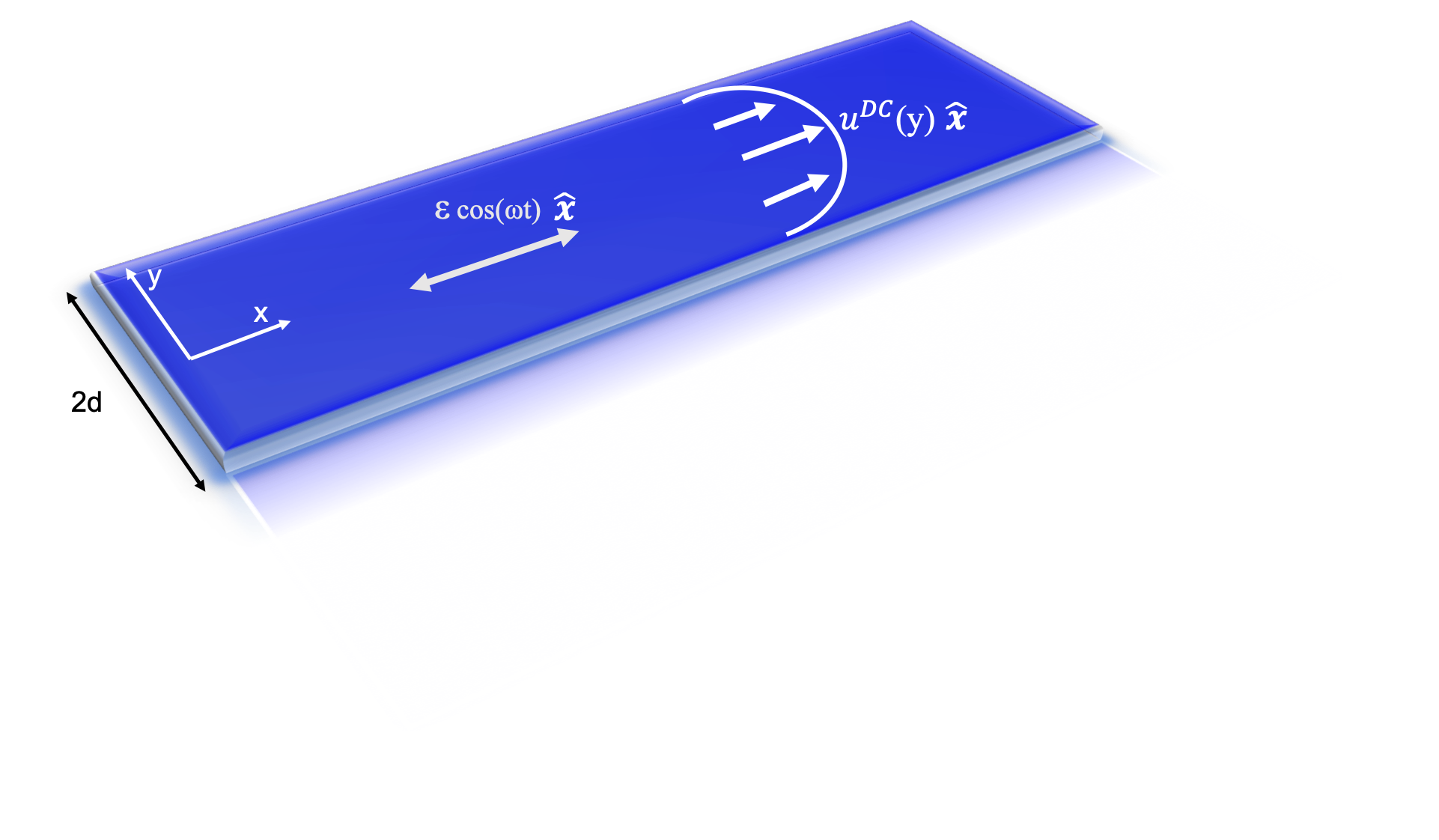}
\vspace{-80pt}
\caption{Flow of an electron liquid in a Hall bar driven by a uniform AC electric field in the $\hat{\bm{x}}$ direction. For nonreciprocal liquids, a  DC electric current appears that is quadratic in the amplitude of the AC electric field. 
\label{hall_bar}}
\end{figure}

Nonlinear current rectification has been experimentally demonstrated in graphene devices for various frequencies of the AC drive  across the diffusive-to-hydrodynamic crossover \cite{Ganichev2022}. 
The required spatial symmetry breaking was artificially engineered either by gating the system in a specific manner or by patterning devices with an asymmetric array of antidots \cite{Yahniuk2024,Hild2024}. Both approaches can be regarded as extrinsic mechanisms of spatial symmetry breaking. On the experimental front, we should also highlight recent photoconductivity measurements of hydrodynamic electrons in graphene \cite{Gallagher2019,Kravtsov2025}, which reveal pronounced viscous effects and complex dependencies on radiation power.

In this work, we explore an intrinsic mechanism of hydrodynamic current rectification in which the required breaking of inversion and TR symmetry arises from the electron liquid itself, rather than device geometry. This distinguishes our treatment from earlier hydrodynamic approaches to ratchet-like effects in spatially modulated 2D systems \cite{Shur2015,Kachorovskii2025}.   
To isolate the effects of intrinsic nonreciprocity of the electron liquid, we consider AC-driven flows in the Poiseuille geometry. In this geometry, the nonlinear terms present in the standard Navier-Stokes equation do not produce current rectification. 

The paper is organized as follows. In Sec.~\ref{sec:clean} we consider the nonlinear hydrodynamic effect in the absence of disorder.
In Sec.~\ref{sec:main_equations} we obtain the generalization of the Navier-Stokes equation governing time-dependent flows of nonreciprocal liquids. In Sec.~\ref{sec: response} we use it to evaluate the DC current for a monochromatic drive of frequency $\omega$. 
In Sec.~\ref{sec: memory} we consider the oscillatory part of the nonlinear current and show that the hysteretic current-voltage dependence is significantly affected by the nonreciprocity of the electron liquid. In section \ref{sec: disorder}
we consider the effect of weak disorder-induced friction on the DC current. Finally, in section \ref{sec: TI} we 
consider the current induced by an instantaneous change of the bias voltage from zero to a steady-state value. Because of nonlinearity, the saturation of current to its steady state value in the transient region is not described by an exponential time-dependence. The main presentation is accompanied by several technical appendices that provide additional details of the calculations. 

\section{Nonlinear Poiseuille flow of nonreciprocal liquids driven by \emph{ac} electric fields}
\label{sec:clean}

\subsection{Main equations and scales}  
\label{sec:main_equations}

We consider a two-dimensional (2D) 
flow of an electron liquid in the Hall-bar geometry shown in Fig. \ref{hall_bar}. 
In the hydrodynamic regime, the evolution equation for the momentum density $\bm{p}=\rho\bm{v}$ of the electron system with mass density $\rho$ takes the form of Newton’s second law:
 \begin{equation}
 \frac{\partial \bm{p}}{\partial t}=-\bm{\nabla}\cdot\hat{\bm{\Pi}}+en\bm{E},
 \end{equation}
where $n$ is the density of electrons, $e$ is the electron charge, $\bm{E}$ is the electric field. The momentum flux tensor density, 
\begin{equation}
\hat{\bm{\Pi}}\equiv\Pi_{ij}=P\delta_{ij}+\rho v_iv_j-\sigma_{ij},
\end{equation}
comprises of the local pressure term $P\delta_{ij}$ and the equilibrium part of the momentum flux tensor $\rho v_iv_j$
\footnote{We note that in systems lacking Galilean invariance
the latter should be viewed as the lowest order term in a formal expansion of momentum flux density in an equilibrium state of uniform flow to second order in the powers of the flow velocity $\bm{v}$. In the following we omit the possible higher order terms. }.  

The dissipative part of the momentum flux is described by the Cauchy stress tensor, which is proportional to the gradients 
of the velocity 
\begin{equation}
\sigma_{ij}=2\eta V_{ij}+\delta\sigma_{ij}, \quad V_{ij}=\frac{1}{2}(\partial_jv_i+\partial_iv_j), 
\end{equation}   
where $V_{ij}$ is the strain rate tensor and $\eta$ is the shear viscosity of the fluid. For simplicity, we consider incompressible liquids and neglect the bulk viscosity. 

The nonreciprical correction  $\delta\sigma_{ij}$ to the Cauchy stress tensor arises only when both TRS and inversion symmetry are broken. It has the general form, cf. \cite{Kirkinis2025}
\begin{align}
    \label{eq:delta_sigma}
    \delta \sigma_{ij} = \eta N_{ijklm}v_k \partial_lv_{m}.
\end{align}
The fifth-rank tensor $N_{ijklm}$ characterizes the degree of nonreciprocity of the liquid and has the physical dimension of the inverse velocity. Its nonzero value requires breaking of inversion symmetry and TRS, which may be caused by an external magnetic field or spontaneous magnetic ordering in the liquid. The tensor structure  of $N_{ijklm}$ depends on the lowering of rotation symmetry induced by TRS breaking~\cite{Kirkinis2025}.  The nonreciprocal stress $\delta \sigma_{ij}$ may be induced, for instance, by I) application of an in-plane magnetic field to an electron liquid with spin-orbit coupling, or II) valley polarization in graphene.  Case I) and case II) correspond, respectively, to a vector- and tensor- type lowering of rotation symmetry in the terminology of Ref.~\cite{Kirkinis2025}. 

For a current flowing through a channel of width $2d$, as displayed in Fig. \ref{hall_bar}, the velocity field has only one nonzero component $\bm{v}(\bm{r},t)=u(y,t)\hat{\mathbf{x}}$, cf. \cite{Landau1987}, which depends on the transverse coordinate of the Hall-bar. Thus, only the ${xy}$ component of the stress tensor is nonzero. Its nonreciprocal part
\be
\delta \sigma_{xy} = \eta{N} u \partial_y u
\ee
is characterized by a single parameter $N$, which depends on the form of rotational symmetry breaking~\footnote{
The breaking of continuous rotation symmetry that is required for the appearance of the nonreciprocal correction to the viscous stress tensor in Eq.~\eqref{eq:delta_sigma}, also produces an antisymmetric viscous stress that is proportional to the flow vorticity. The corresponding rotational viscosity was considered for instance in Refs. \cite{Zahn1995,Rinaldi2002,Rinaldi2002thesis,Chaves2008,Doornenbal2019}. Since in the Poiseuille flow the vorticity reduces to $ -\partial_y u$, this contribution is already included in Eq.~\eqref{ns0} provided $\eta$ is a linear combination of the shear and rotational viscosity. We work to lowest order in rotational symmetry breaking. Within this accuracy $\eta$ is equal to the shear viscosity.}.
For vector type of symmetry breaking, caused by an in-plane magnetic field $\bm{B}$, $N$ depends on the angle between $\bm{B}$ and the $\hat{\bf x}$-axis. For tensor-type of symmetry breaking that corresponds to lowering the rotation symmetry to $C_3$, $N$ depends on the angle $\theta$ between $\hat{\bf x}$ and the principal directions of the $C_3$ group. In these two cases $N$ takes the form, cf. \cite{Kirkinis2025}
\begin{equation}\label{N}
{N} = \left\{\begin{array}{cc} \alpha B, & \text{vector-type}, \\ |r|\cos(3\theta), & \text{tensor-type}.\end{array}\right.
\end{equation}
The parameter $\alpha$ depends on the electron band-structure, the strength of spin-orbit coupling and the field orientation.
We note that nonreciprocity of the tensor-type is realized in the recently discovered quarter-metal state in rhombohedral trilayer graphene~\cite{Zhou2021}. In  this system the breaking of TRS is caused by spontaneous valley polarization and the lowering of the rotation symmetry to $C_3$ is caused by the trigonal warping of the electron spectrum. The parameter $|r|$ may be estimated in this case as $\Delta/v_F\epsilon_F$, where $\Delta$ is the energy of trigonal warping as discussed in \eqref{eq:trigonalwarpingspectrum}.  Angle-resolved transport reveals pronounced nonreciprocity due to spontaneous symmetry breaking in twisted trilayer graphene \cite{Zhang2024}. In this case, the angle $\theta$ depends on the relative orientation between the graphene lattice and the direction of flow in the channel 
The equation of motion for the fluid element in the presence of a spatially uniform time-dependent electric field along the channel reduces to  
\begin{equation} \label{ns0}
\rho\partial_tu=en\mathcal{E}(t)+\eta\partial_y[\left(1+{N}u \right)\partial_yu], 
\end{equation}
where we absorbed the pressure gradient $\partial_xP$ into the electromotive force (EMF) density $en\mathcal{E}=enE-\partial_xP$. 

\subsection{\label{sec: response}Monochromatic drive} 

In what follows we will consider a monochromatic EMF
\be\label{E}
\mathcal{E}(t)=\mathcal{E}\cos(\omega t)
\ee
of amplitude $\mathcal{E}$ and frequency $\omega$ driving the electron liquid in the $\hat{\bm{x}}$ direction of a Hall-bar of width $2d$, as shown in figure \ref{hall_bar}. As we will show below, in addition to the oscillatory current, the flow of nonreciprocal liquids contains a net DC current which is quadratic in the driving field $\mathcal{E}$.

\begin{figure}
\includegraphics[width=\linewidth]{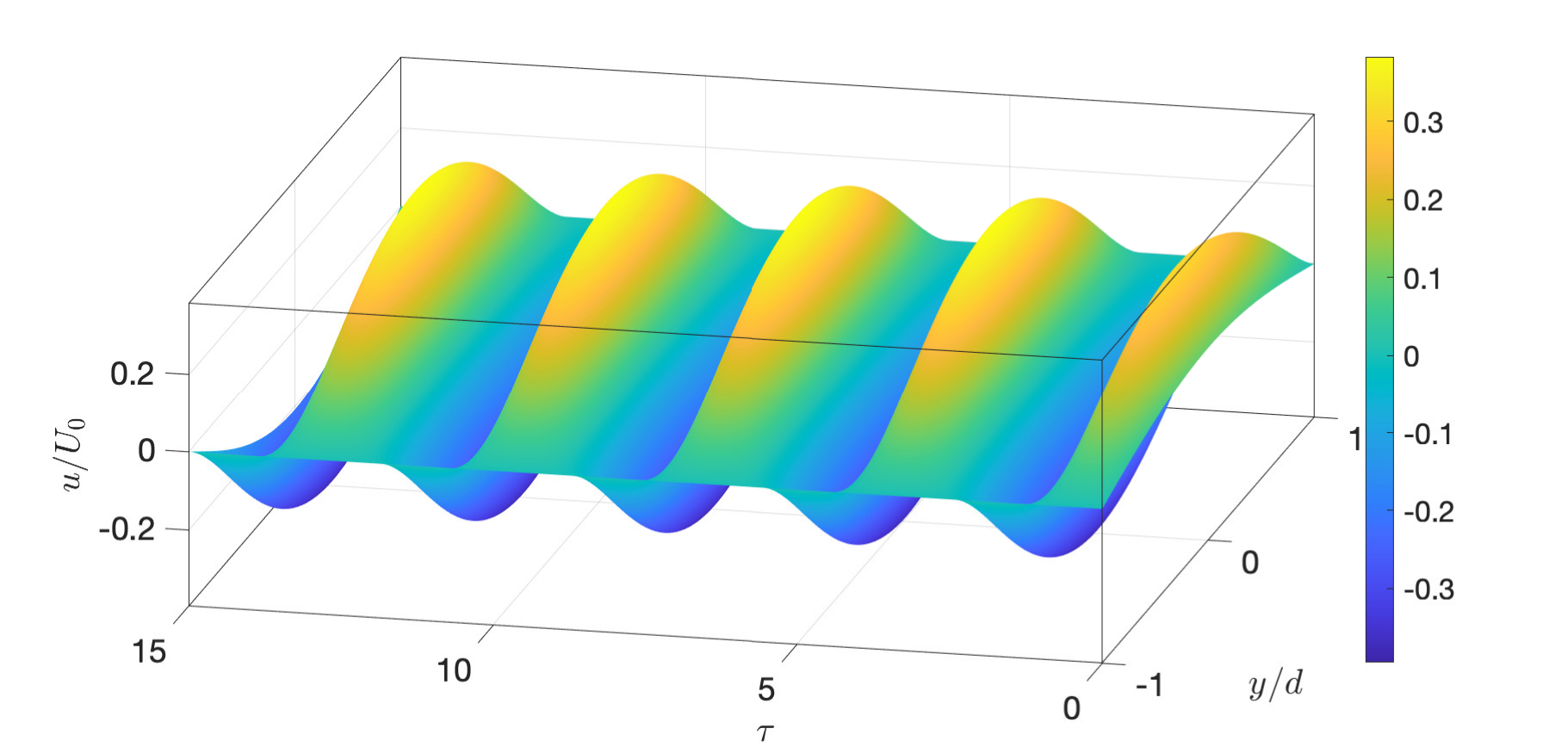}
\caption{Characteristic scaled velocity $u(y,\tau)/U_0$ profiles obtained by numerical solution of the time-dependent Stokes Eq. \eqref{ut0} (with oscillating electric field) that we average over time and space to obtain the current denoted by the circle markers in panels (a) and (b) of figure \ref{I_Re_num}. 
\label{uytfig}}
\end{figure}

To reduce the problem to dimensionless variables, we introduce the characteristic scales $U_0$ and $I_0$ which describe the hydrodynamic velocity (averaged over the width $2d$ of the channel) and the net current of a flow driven by a static field of amplitude $\mathcal{E}$ 
\begin{equation}\label{U0N}
U_0 = \frac{en\mathcal{E}d^2}{\eta}, \quad
I_0 = \frac{2}{3}  \frac{(en)^2\mathcal{E}d^3}{\eta}.
\end{equation}
We define the dimensionless nonreciprocity number $\mathcal{N}$ and the vibrational number $\mathcal{R}$,
\begin{equation}\label{NRev}
\mathcal{N} = N U_0,\quad 
\mathcal{R}= \frac{\omega d^2}{\nu},
\end{equation}
where $\nu = \eta/\rho$ is the kinematic viscosity. 
We also introduce
dimensionless variables
\begin{equation} \label{ytus}
y\rightarrow \frac{y}{d}, \quad \tau=\omega t,\quad u \rightarrow \frac{u}{U_0},
\end{equation}
so that $y\in(-1,1)$. The momentum equation becomes
\begin{equation}\label{ut0}
\mathcal{R}\partial_\tau u = \cos\tau + \partial_y \left[ (1+ \mathcal{N} u)\partial_y u \right]. 
\end{equation}
A characteristic numerical solution $u(y,t)$ of \rr{ut0} is displayed in Fig. \ref{uytfig}. 
In the presence of the oscillating field \rr{E}, the flow acquires a new symmetry, in comparison to the one present in the static case \cite{Kirkinis2025},
whereby the change of sign 
$N\to -N$ (possible by changing the magnetic field direction in the vector case, for instance) changes the sign of the DC current.

\begin{figure*}[t!]
\includegraphics[height=2.7in,width=7.4in]{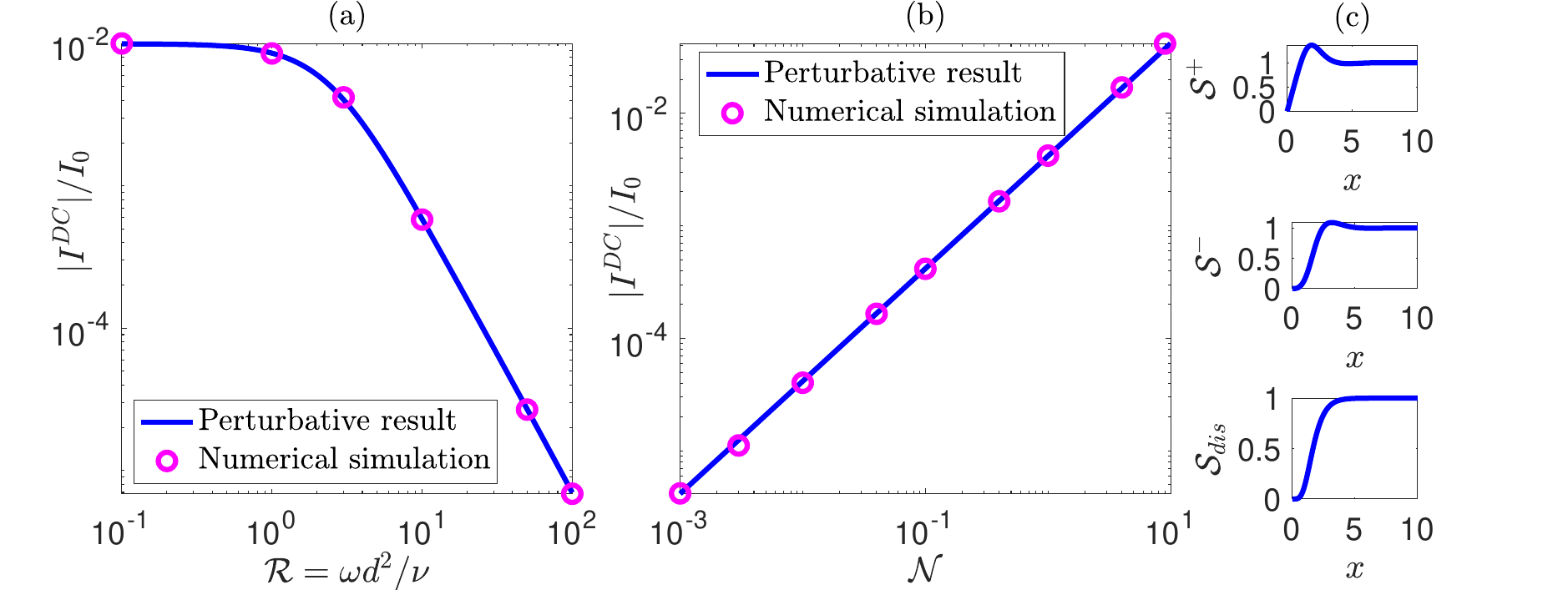}
\caption{
Panel (a) continuous curve: 
Absolute value of scaled DC current \eqref{IavReN} vs. the vibrational number $\mathcal{R}= \omega d^2/\nu$ for the nonreciprocal number $\mathcal{N} =0.1$. At low frequencies
it reaches the value $\mathcal{N}/10$, as predicted by the first term in the series expansion in \eqref{IavRe}. At high frequencies it decays as $\mathcal{R}^{-2}$ as predicted in the asymptotic expansion of Eq. \eqref{IavRe}. 
Panel (b) continuous curve: Absolute value of scaled DC current \eqref{IavReN} vs. the nonreciprocal number $\mathcal{N}$ for the experimentally relevant magnitude of the vibrational number $\mathcal{R} =12$. 
Circle markers denote scaled averaged current $|I|$ obtained by solving numerically the time-dependent Stokes Eq. \eqref{ut0}. There is good agreement between the perturbative result \eqref{IavRe} and the full numerical solution of the momentum equation \eqref{ut0}. 
Panels (c): structure factors defined in \rr{Spm} and \rr{Sdis}. 
\label{I_Re_num}}
\end{figure*}

Equation \rr{ut0} can be solved perturbatively with respect to a small nonreciprocity number $\mathcal{N}\ll 1$. However, as it will be shown below, the perturbative result provides a good representation for the observables even at values $\mathcal{N} \sim 1$. 
We express the velocity field in the form
\begin{align}
u(y,t) = \Re \!\left\{u_0^{AC}(y)e^{-i\omega t} + \!\mathcal{N}\!\! \right.& \left. \left[ u_1^{DC}(y) +  u_1^{AC}(y)e^{-2i\omega t}  \right]\!\right\}\nonumber \\ 
& + O(\mathcal{N}^2) 
\label{uyt}
\end{align}
where $u_0^{AC}$ and $u_1^{AC}$ are complex fields and $u_1^{DC}$ is real.
Thus, the velocity field $u(y,t)$ is oscillatory (cf. Fig. \ref{uytfig}), but because the momentum equation is quadratic in $u$, it also consists of a zero (DC) mode \citep{Kirkinis2014zero,Shrestha2025,Shrestha2025b}. 

At $\mathcal{N}\equiv 0$ one obtains the velocity profile of the well-known oscillatory pressure-driven flow, cf. \cite{Landau1987} and Appendix \ref{sec: appendixPressure}. In dimensional variables the zeroth order velocity $u_0 = \Re\left\{u_0^{AC}(y)e^{-i\omega t} \right\}$, satisfying no-slip boundary conditions at $y = \pm d$, becomes 
\begin{equation} \label{uLandau0}
u_0 = \Re \left\{ \frac{-en\mathcal{E}}{\eta k^2} e^{-i\omega t} \left[ 1 - \frac{\cos k y}{\cos kd} \right]   \right\}
\end{equation}
where $\Re$ denotes the real part of a complex expression and
\begin{equation}\label{kdelta}
i\omega = \nu k^2, \quad  k = \frac{1+i}{\delta}, \quad \delta = \sqrt{\frac{2\nu}{\omega}},
\end{equation}
is the familiar notation of the penetration depth $\delta$ of the Stokes boundary layer, and complex wavenumber $k$ of the oscillating pressure-driven flow in a channel, cf. \cite[p.84-89]{Landau1987} and Appendix \ref{sec: appendixPressure}. 

It is easy to determine the form of the $O(\mathcal{N})$ corrections to the velocity in Eq. \rr{uyt}, see Appendix \ref{sec: u1ACDC} for the derivation. 
The DC current is then $O(\mathcal{N})$ and has the form
\begin{equation} \label{IavReN}
I^{DC}= I_0 \frac{3\mathcal{N}}{8} \left[
-\frac{2}{\mathcal{R}^{2}}+\frac{ \sqrt{2}}{\mathcal{R}^{\frac{5}{2}}}\mathcal{S}^{+}(\sqrt{2\mathcal{R}})
\right].
\end{equation}
Here, and in what follows, we employ the structure factors
\be \label{Spm}
\mathcal{S}^{\pm}(x) = 
\frac{\sinh x\pm\sin x}{ \cosh x+\cos x}, 
\ee
which are displayed in panel (c) of Fig. \ref{I_Re_num}. 
For small and large vibrational numbers 
Eq.~\eqref{IavReN} becomes
\be
\label{IavRe}
I^{DC} \approx I_0 \mathcal{N} 
\left\{
\begin{array}{cc}
\displaystyle
-\frac{1}{10}+\frac{31 \mathcal{R}^{2}}{1890}+\ldots &\textrm{as } \mathcal{R}\ll1, \\
\displaystyle
-\frac{3}{4\mathcal{R}^2}+\frac{3\sqrt{2}}{8\mathcal{R}^{5/2}} +\ldots& \textrm{as } \mathcal{R}\gg1.
\end{array}
\right.
\ee

Restoring the physical dimensions in Eq.~\eqref{IavRe} we express the magnitude of the DC current in terms of the EMF $\mathcal{E}$ and the parameters of the system
for high and low frequencies of the AC drive.  At low frequencies, $\rho \omega d^2/\eta\ll 1$ it becomes
\be \label{IDCdim1}
I^{DC} \approx 
-N \frac{(en)^3 \mathcal{E}^2d^5}{15\eta^2}. 
\ee
The DC current can also be expressed with respect to a nonlinear conductance
\be \label{G2}
G_2 = -N \frac{(en)^3d^5}{15(\eta L)^2},
\ee
whereby $I^{DC} = G_2 V^2$, $V=\mathcal{E}L$ is the amplitude of the oscillating voltage and $L$ is the length of the Hall-bar. The key point to highlight in this result is that in the low frequencies of the AC drive $I^{DC}$ is super-extensive, namely $I^{DC} \propto d^5$, which is a consequence of nonlocality of hydrodynamic transport \cite{Levitov2016}. 

In the high-frequency regime, 
$\rho \omega d^2/\eta\gg 1$
it becomes 
\be\label{IDCdim2}
I^{DC} \approx 
-N \frac{(en)^3 \mathcal{E}^2d}{2(\rho \omega)^2}. 
\ee

In panel (a) of figure \ref{I_Re_num} with the continuous curve we display the absolute value of the perturbative DC current \eqref{IavRe} as a function of the vibrational number for the nonreciprocal number $\mathcal{N} =0.1$. The asymptotics present in expressions \eqref{IavRe} are clearly visible, that is, for small $\mathcal{R}$ the current reaches the plateau $\mathcal{N}/10 = 10^{-2}$ and for large $\mathcal{R}$ the line has slope $-2$ (the current decays as $\mathcal{R}^{-2}$).  

In panel (b) of figure \ref{I_Re_num} with the continuous curve we display the absolute value of the perturbative DC current \eqref{IavRe} as a function of the nonreciprocal number for the experimentally-relevant value of the vibrational number $\mathcal{R} =12$ (in the terahertz frequencies, see discussion below). The curves are straight lines of slope $1$, showing that the perturbative result is valid even at relatively high nonreciprocity number (which is the perturbing parameter).  

As a useful validity check, it is easy to determine the adiabatic limit of the oscillating net current $I =\int_{-d}^d enu(y) dy$ (where $u$ is given by \rr{uyt}, cf. Appendix \ref{sec: u1ACDC}), which reads
\begin{equation} \label{IfNser}
I =I_0\left[ 1- \frac{\mathcal{N}}{5} + O(\omega^2) \right] 
\end{equation} 
i.e., it is the equilibrium current derived in \cite[Eq. (11) \& (12) and Fig. 2]{Kirkinis2025} (see also Eq. \eqref{Iss} and \rr{fN0}) expanded for small $\mathcal{N}$. Thus, whereas at equilibrium one obtains a correction to the current
$\delta I_{hydro} = -\mathcal{N}/5$, here the corresponding correction \rr{IavReN} is frequency dependent. A comparison of the two currents in the adiabatic limit gives $I^{DC} =\delta I_{hydro}/2 $ and they are of the same order of magnitude. 

To verify the validity of the above statements, we solve the time-dependent Eq. \eqref{ut0} numerically starting with an initial condition $u(y,0) = \frac{1}{10}\cos(\pi y /2)$, and form the current $ I = \frac{1}{T}\int_0^T d\tau \int_{-d}^d enu(y,\tau) dy$ by spatially integrating the numerical solution (see Fig. \ref{uytfig}) and numerically averaging over the period of oscillation $T$. The period is the interval between two maxima determined after the effect of initial conditions has elapsed and a steady oscillatory state is established. The frequency of oscillation is then nearly equal to the one of the external forcing. Circle markers in panels (a) and (b) of figure \ref{I_Re_num} denote the scaled averaged current $|I|$ at the indicated Reynolds numbers for  the nonreciprocal number $\mathcal{N} =0.1$. There is good agreement between the perturbative result \eqref{IavReN} (continuous curve in panels (a) and (b) of Fig. \ref{I_Re_num}) and the full numerical solution of the time-dependent Stokes equations \eqref{ut0} (circle markers, same panels). 

The asymptotic expansion in \rr{IavRe}, Eq. \rr{IfNser} and figure \ref{I_Re_num} can be employed to compare the magnitude of nonreciprocity at experimentally-relevant high frequencies, to that in the disordered-dominated regime
adopting the arguments followed in the non-oscillating nonreciprocal case \cite[Eq. (17)]{Kirkinis2025}.
From figure \ref{I_Re_num} and employing the notation of length scales for electron-impurity ($\ell_{ie}$), and electron-electron ($\ell_{ee}$) mean free paths, and disordered current $\delta I_{dis}$ introduced in \cite[Eq. (15)]{Kirkinis2025}, we conclude that 
\be \label{relcor}
\frac{I^{DC}}{\delta I_{dis}} \sim \frac{1}{20} \frac{d^2}{\ell_{ee}\ell_{ei}}
\ee
where, by considering an experimental estimate for the kinematic viscosity $\nu\sim 0.1$ m$^2$/s \cite{Bandurin2016,Zeng2024}, Hall-bar width $d\sim 1$ $\mu$m and a frequency of the order of one terahertz, we obtain that $\mathcal{R}=10$ and from figure \ref{I_Re_num} this gives 
$I^{DC}/I_0\sim 10^{-3}$. Therefore, the relative correction \rr{relcor} is still enhanced by a large factor and this is analogous to, but not identical to, the enhancement relative to the non-oscillating nonreciprocal case, cf. \cite[Eq. (17)]{Kirkinis2025}. 

In conclusion to this part of the analysis, we note that the momentum density can, in principle, contain also a contribution that is quadratic in the velocity of the liquid, which corresponds to nonreciprocal inertia of the liquid (for its derivation see Appendix \ref{sec: v2}). However, this contribution does not lead to a zero-mode (DC) current, which is the main result of this paper, but only affects the oscillatory velocity field. We justify these arguments in Appendix \ref{sec: v2}.

\begin{figure}[t!]
\includegraphics[width=\linewidth]{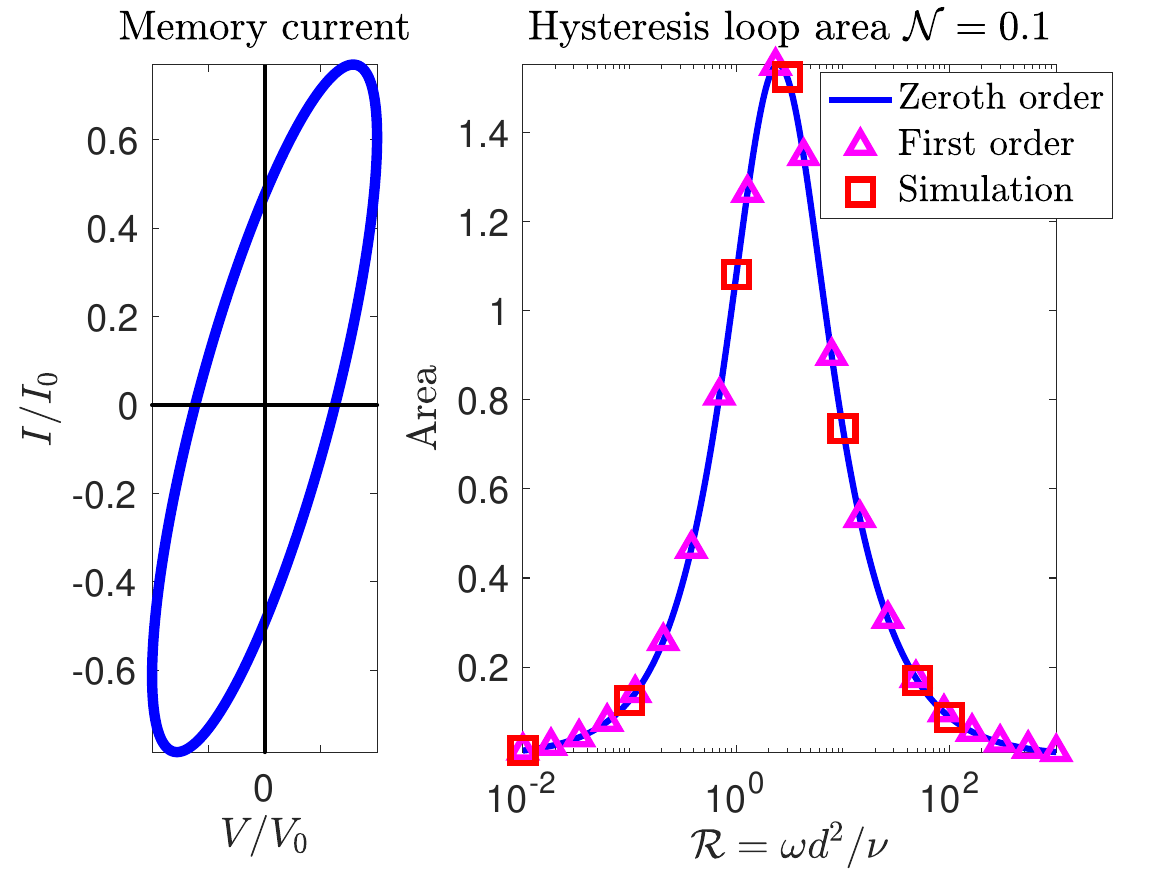}
\caption{Left panel: Current-voltage hysteresis loop evaluated at $\mathcal{R}=2$ which maximizes the loop area (i.e. maximum energy dissipation, cf. the $\mathcal{R}$ at the area maximum in the right panel). Right panel, continuous curve: Area between zeroth order current \eqref{I0ACb} and voltage. The triangle markers denote the area formed between the current \eqref{I0ACb} that includes first order corrections. The square markers denote the area between current and voltage by running numerical simulations of the 
full momentum equation \eqref{ut0}. The frequency that maximizes the area of the loop, determines the characteristic `memory' time $\tau_M \sim \frac{d^2}{2\nu}$, cf. \cite[Eq.(3) \& Fig. 3C]{Robin2023} and  \cite[Fig. 3(c)]{Kamsma2023PRL}. 
\label{memory1}}
\end{figure}

\subsection{\label{sec: memory}Memory effects}

The strong dependence of the nonlinear DC current on the frequency of the driving field points to the importance of retardation and memory effects in the hydrodynamic photogalvanic effect. Conversely, nonlinearity also affects the retardation of the oscillatory part of the current with respect to the driving voltage. This situation is reminiscent of circuit elements that exhibit memory, that is, displaying properties depending on system state and history, as they were introduced in the work of Chua \& Kang \cite{Chua1976}. 

The hysteretic dependence of the net current on the instantaneous electric field in the presence of retardation
is typically represented by Lissajous or Bowditch figures. For the AC-driven nonreciprocal flows, the shape of the Lissajous curves depends not only on the frequency, but also on the amplitude of the drive.  
We thus proceed to examine the effects of nonreciprocity on the shape of these curves. On the left panel of figure \ref{memory1} we display the net instantaneous current $I(\tau)/I_0$ calculated by numerical solution of Eq.~\eqref{ut0}, vs. the instantaneous voltage $V(\tau)/V_0 = \cos \tau$, where $V_0 = \mathcal{E}L$, for fixed nonreciprocal number $\mathcal{N} = 0.1$ and $\mathcal{R} =2$ (which maximizes the loop area as displayed in the right panel of the same figure).  

\begin{figure}[t!]
\includegraphics[width=\linewidth]{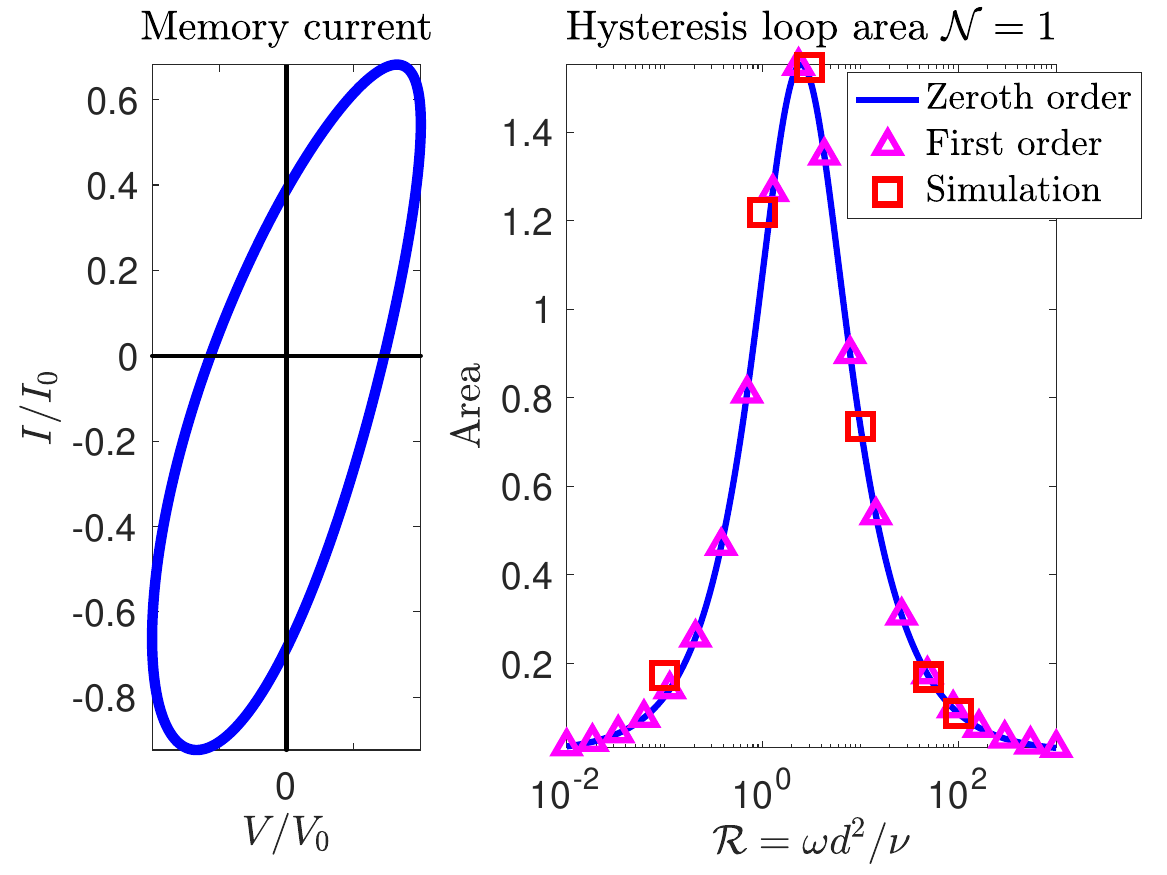}
\caption{Same as in figure \ref{memory1} but for $\mathcal{N} =1$. New features include an increased skewness of the hysteresis loop as $\mathcal{N}$ increases (compare left panel with its counterpart of Fig. \ref{memory1}). The hysteresis loop area also increases with $\mathcal{N}$. The approximate current (continuous line and triangle markers) starts to deteriorate. \label{memory3}}
\end{figure}

On the other hand, as seen in the left panel of figure \ref{memory3}, for larger nonreciprocal number $\mathcal{N} =1$ there is a qualitative change in the shape of the hysteresis loops which become skewed followed by a corresponding area increase.

We compare the numerical results denoted by the square markers in the right panels of figures \ref{memory1} and \ref{memory3} with the analytic predictions of the perturbative results. To this end, 
the continuous curve in the right panels of figures \ref{memory1} and \ref{memory3} is the area formed between the zeroth order current obtained from \eqref{uLandau0}
\begin{align}
\label{I0ACb}
\frac{I^{(0)AC}(\tau, \mathcal{R})}{I_0} =& \frac{3}{\sqrt{2}}\left\{ \left[ \frac{\sqrt{2}}{\mathcal{R}}
-\frac{1}{\mathcal{R}^{\frac{3}{2}}}
\mathcal{S}^{+}(\sqrt{2\mathcal{R}})
\right]\sin \tau \right.\nonumber \\ 
&  \left. + \frac{1}{\mathcal{R}^{\frac{3}{2}}}
\mathcal{S}^{-}(\sqrt{2\mathcal{R}})\cos \tau \right\}
\end{align}
and voltage $V(\tau)/V_0 = \cos \tau $ for the designated vibrational and nonreciprocity numbers, whereby the structure factors $\mathcal{S}^\pm$ were defined in \rr{Spm}.
The triangle markers denote the area formed between the current and voltage that includes first order corrections. There is good agreement between theory and simulation. 

Although memresistive circuits appear with pinched hysteresis loops (the current vanishes when the voltage does), memcapacitive systems display charge-voltage loops that are not pinched; cf.
\cite{Krems2010}, \cite[\S 4.2.2]{Pershin2011} for the ionic case. Such open hysteresis loops are also observed in the experiments of \cite[Fig. S13]{Kamsma2023}. In addition, co-existence of memresistive and memcapacitive behavior has been observed in experiments \cite{Liu2006,Salaoru2013}. See also the discussion in \cite[\S\S 2.7 \& 4.1.3]{Pershin2011}. 

Following these contributions we can associate with ``memory'' the inverse frequency giving the largest area in an $I(V )$ hysteresis loop.
We define a memory time-scale $\tau_M$ to be proportional to the largest loop area, which takes place when the vibrational number is about $2$. Thus, for kinematic viscosity $\nu = 10^{-1} \textrm{ m}^2/$sec and Hall-bar width $d = 10^{-6}$ m, we obtain a memory time $ \tau_M \sim 10^{-11}$ sec. 

\begin{figure*}[t!]
\vspace{5pt}
\begin{center}
\includegraphics[height=3in,width=6in]{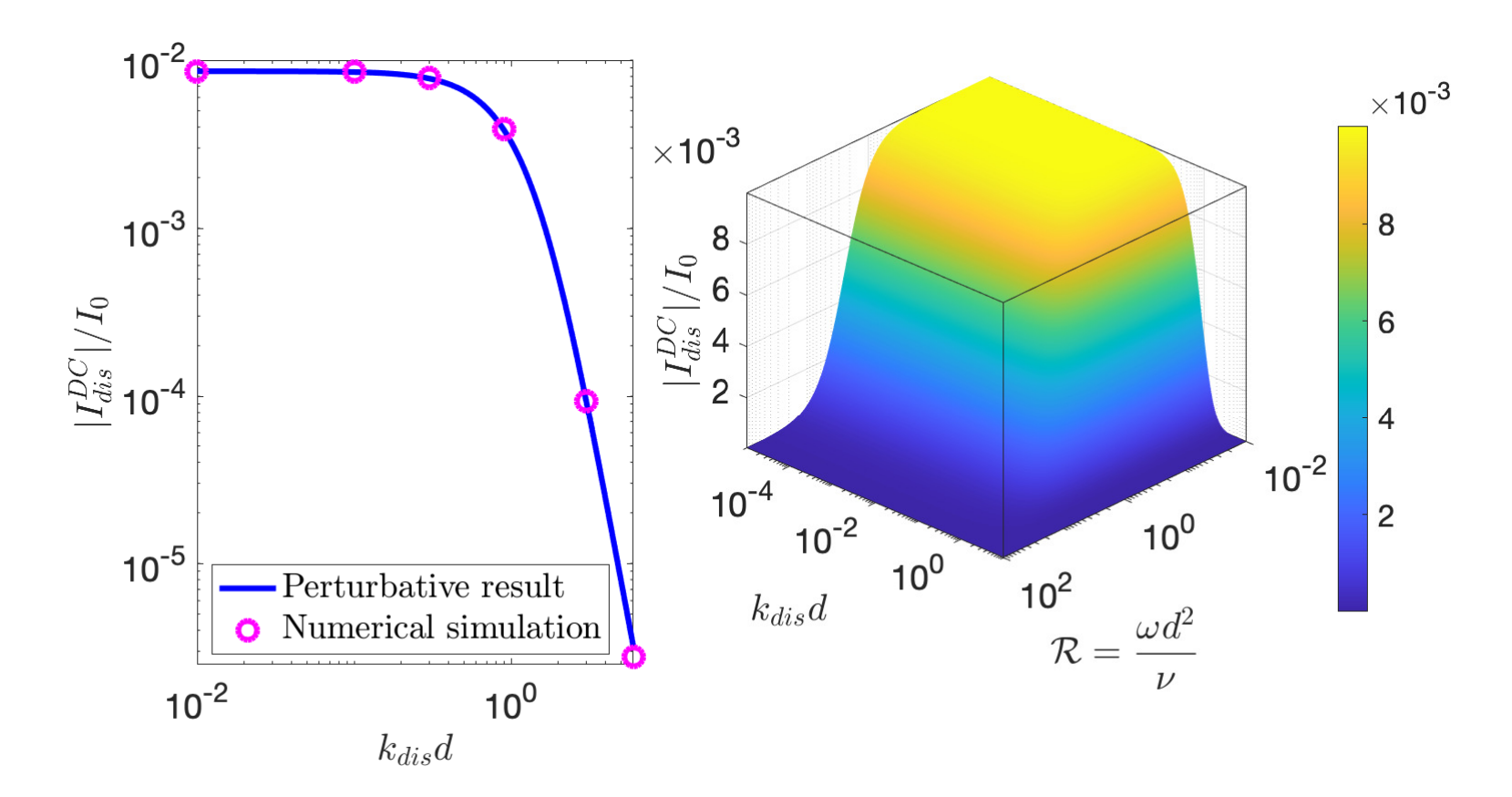}
\vspace{-0pt}
\end{center}
\caption{Left panel: absolute value of the scaled perturbative in $\mathcal{N}$ DC current \rr{IDCdis} vs. the dimensionless number $k_{dis} d$ for the nonreciprocal number $\mathcal{N} =0.1$. There is good agreement between the perturbative result \rr{IDCdis} and full numerical simulations of \rr{NSdis} for all values. For small disorder or narrow channels the current converges to the plateau $I^{DC}/I_0 = 10^{-2}\equiv \mathcal{N}/10$. For large disorder or wide channels the current decays as $(dk_{dis})^{-5}$ (the straight line has slope equal to $-5$), agreeing with the asymptotic result \rr{asymptdis}. 
Right panel: as in the left but now also vary the vibrational number $\mathcal{R}$. The plateau at $I^{DC}/I_0 = 10^{-2}\equiv \mathcal{N}/10$ is the same as in the clean case as predicted by the first term in the expansion \rr{IavRe}, 
cf. Fig. \ref{I_Re_num}. 
\label{IDC_dis}} 
\vspace{-0pt}
\end{figure*}

\section{\label{sec: disorder}Photogalvanic effect in the presence of weak disorder}

In high mobility samples, where the momentum relaxation rate is much smaller than the rate of electron-electron collisions, the effect of disorder may be taken into account by introducing a phenomenological friction coefficient into the hydrodynamic equations. If the correlation radius of the disorder potential is greater than the length of electron-electron scattering, its value may be obtained by averaging the hydrodynamic equations over the realizations of the disorder \cite{Andreev2011,Li2020,Lucas2016,Li2022}. 

In this section we consider the effect of disorder on the photogalvanic current. The momentum equation \rr{ut0} acquires a disorder-induced friction force that is proportional to the velocity field of the electron liquid
\be \label{NSdis}
\mathcal{R} \partial_\tau u = \cos\tau + \partial_y \left[ (1+ \mathcal{N} u)\partial_y u                 \right] - (k_{dis} d)^2 u.
\ee
Here the wavenumber $k_{dis}$ 
is the inverse of the Gurzhi length $l_G$ \cite{Gurzhi1963}.
In principle, $k_{dis}$ also depends on the frequency of oscillation $\omega$. However, it can be shown that in the limit of small frequencies the wavenumber retains its constant form to leading order in $\omega$. We adopt this limit in the discussion that follows. 

On dimensional grounds, the averaged over the period of oscillation net electric current
is expected to have the self-similar form 
$
I^{DC}_{dis}=I^{DC}_{dis} \left(\mathcal{R}, \mathcal{N}, k_{dis}d\right)$.  
In the presence of disorder, the zeroth-order velocity field \rr{uLandau0} remains unchanged by effecting the transformation $k\rightarrow K$, whereby
\be\label{K}
K = \sqrt{\frac{i\omega}{\nu} -  k^2_{dis} }
\ee
and higher order corrections are discussed in Appendix \ref{sec: appendixdis}.
The net disordered DC current \rr{IDCdis}
leads to the relatively long expression \rr{IDCdis2} that we relegate to Appendix \ref{sec: appendixdis}. The adiabatic limit of the disordered DC current \rr{IDCdis2} is 
\be  \label{IDCdisR0}
I^{DC}_{dis} = 
-\frac{\mathcal{N} I_0}{4 (dk_{dis})^{5}}
\mathcal{S}_{dis}(k_{dis} d)
\quad \textrm{as} \quad \mathcal{R}\rightarrow 0,
\ee
where $\mathcal{S}_{dis}$ is the structure factor
\be \label{Sdis}
\mathcal{S}_{dis}(x) = \left[1+2\sech^2 x\right] \tanh x - 3x\sech^2 x
\ee
that we display in panel (c) of Fig. \ref{I_Re_num}. 
The large $k_{dis}$ asymptotics of the current read 
\be \label{asymptdis}
I^{DC}_{dis} = 
-\frac{\mathcal{N}I_0}{4 (d k_{dis})^{5}}+O\! \left((dk_{dis})^{-9}\right)\quad \textrm{as} \quad d k_{dis}\gg 1. 
\ee
In the small disorder limit one recovers for instance the clean result \rr{IavRe} as $\mathcal{R} \rightarrow 0$ and thus
the current converges to the plateau $I^{DC}/I_0 = \mathcal{N}/10$ as predicted by the first term in the series expansion \rr{IavRe}, 
cf. Fig. \ref{I_Re_num}. 

In figure \ref{IDC_dis} we display the averaged over disorder DC current vs. the dimensionless parameters $k_{dis}d$ and $\mathcal{R}$. The perturbative current is compared with the results of full numerical simulations of Eq. \rr{NSdis} in the left panel of figure \ref{IDC_dis}. There is good agreement between the perturbative result \rr{IDCdis} and full numerical simulations of \rr{NSdis} for all values of the parameter $k_{dis}d$. For small disorder or narrow channels the current converges to the plateau $I^{DC}/I_0 = 10^{-2}\equiv \mathcal{N}/10$ which is the same as in the clean case as predicted by the first term in the expansion \rr{IavRe}, 
cf. Fig. \ref{I_Re_num}. For large disorder or wide channels the current decays as $(dk_{dis})^{-5}$ agreeing with the asymptotic result \rr{asymptdis}. 

Finally, we note that Eq. \rr{IDCdisR0}
defines a nonlinear conductance 
\be
G_2^{dis} = -N \frac{(en)^3 d^5}{6(\eta L)^2}
\left\{
\begin{array}{cc}
\displaystyle
\frac{2}{5} &\textrm{as } {dk_{dis}}\ll 1, \\
\displaystyle
(k_{dis} d)^{-5}& \textrm{as } dk_{dis}\gg1.
\end{array}
\right.
\ee
in a two-terminal set-up with DC electric current
\be \label{IGV3}
I^{DC}_{dis}=G_2^{dis} V^2. 
\ee
The weak disorder limit thus recovers Eq. \rr{G2}. 

\begin{figure}[t!]
\includegraphics[height=2.4in,width=3.2in]{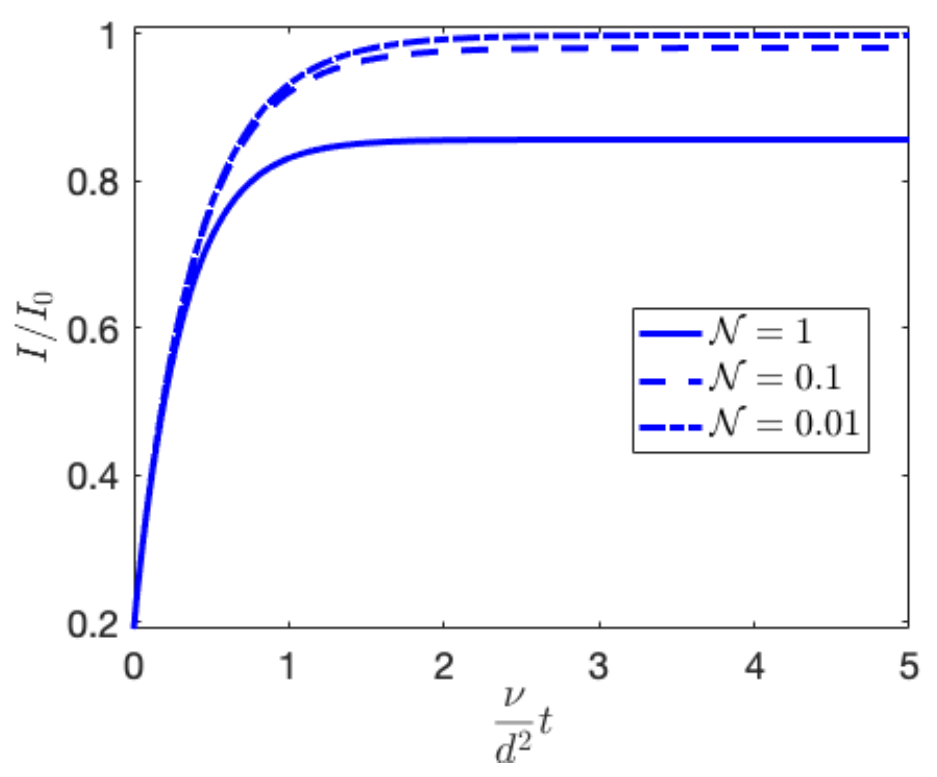}
\caption{Time-variation of the current $ I(\tau) = \int_{-d}^d enu(y,\tau) dy$ by numerically solving the momentum equation \eqref{utss} with a \emph{time-independent} electric field. For each nonreciprocal number $\mathcal{N}$, the current saturates and rapidly converges to the steady-state $f(\mathcal{N})$, cf. Eq. \eqref{Iss}. \label{Itsteady} }
\end{figure}

\section{\label{sec: TI} Time-dependence of current in the transient regime }

In this section, we study the time-dependence of the nonlinear current in the situation where the electric field instantaneously increases from zero to the constant value  $\mathcal{E}$.
Our time-dependent analysis then, enables us to describe the transient regime in which the current increases from zero to the steady-state value studied in Ref.~\cite{Kirkinis2025}.

To this end, consider the Hall-bar configuration of figure \ref{hall_bar} but with a uniform \emph{time-independent} electric field. 
The dimensionless momentum equation becomes
\begin{equation} \label{utss}
 \partial_\tau u = 1 + \partial_y \left[ (1+ \mathcal{N} u)\partial_y u                 \right]
\end{equation}
where $y\in(-1,1)$ and $\tau$ denotes the intrinsic time scale $\tau =  \frac{\nu}{d^2} t$. This equation has equilibrium solutions that depend on a single dimensionless parameter -- the nonreciprocal number $\mathcal{N}$ -- which can be obtained by setting the right-hand side of \eqref{utss} equal to zero, cf. \cite{Kirkinis2025}. In this case, the time-independent current $ I= \int_{-d}^d enu(y) dy$ becomes \cite{Kirkinis2025}
\begin{equation} \label{Iss}
I = I_0 f\left( \mathcal{N} \right),
\end{equation} 
where 
$
f(\mathcal{N})=\frac{3}{2\mathcal{N}^{\frac{3}{2}}}\left[(1+\mathcal{N})\arcsin\sqrt{\frac{\mathcal{N}}{1+\mathcal{N}}}-\sqrt{\mathcal{N}}\right]
$
and 
\be\label{fN0}
f(\mathcal{N})\sim 1 - \frac{\mathcal{N}}{5}, \quad \textrm{as } \mathcal{N} \to 0. 
\ee

We now solve the time-dependent Eq.~\eqref{utss} numerically starting with an initial condition $u(y,0) = \frac{1}{10}\cos(\pi y /2)$, and form the time-dependent current $ I(\tau) = \int_{-d}^d enu(y,\tau) dy$ by spatially averaging the numerical solution. In figure \ref{Itsteady} we display the scaled time-dependent current for various values of the nonreciprocal number $\mathcal{N}$. In all cases the solution converges to the aforementioned steady-state \eqref{Iss}, whose magnitude depends on the nonreciprocity number. Thus, each steady state is obtained by evaluating the function $f$ in \eqref{Iss} at the respective nonreciprocity number displayed in the legend of Fig. \ref{Itsteady}.

\section*{Acknowledgements}

We thank Leonid Golub for useful discussions and comments on the paper. The work of L. B. and A. V. A. was supported by the National Science Foundation (NSF) Grant No. DMR-2424364, the Thouless Institute for Quantum Matter (TIQM), and the College of Arts and Sciences at the University of Washington. The work of A. L. was supported by NSF Grant No. DMR-2452658 and H. I. Romnes Faculty Fellowship provided by the University of Wisconsin-Madison Office of the Vice Chancellor for Research and Graduate Education with funding from the Wisconsin Alumni Research Foundation.

\section*{Data availability} 

All data presented in the figures were generated from analytical expressions derived and defined in the paper. The Mathematica code used to produce the plots will be made available by the authors upon reasonable request.

\appendix
\section{\label{sec: appendixPressure}Zeroth order in $\mathcal{N}$ velocity field}

We describe the zeroth order in $\mathcal{N}$ velocity field \rr{uLandau0}, by repeating the solution method developed in \cite[p.89]{Landau1987}. A harmonic in time pressure gradient inside a rectangular channel gives rise to a flow in the $x$-direction (in our notation) satisfying  the $x$-momentum
equation
\be
\partial_t u = c  e^{-i\omega t} + \nu \partial_z^2 u, 
\ee
where the units of the scalar $c$ are implied. 
The solution, satisfying no-slip boundary conditions at $z = \pm d$ is 
\be \label{uLandau}
u = \frac{ic}{\omega} e^{-i\omega t} \left[ 1 - \frac{\cos k z}{\cos kd} \right]
\ee
where $k^2\nu  = i\omega$, as considered in this article. 
Its average value is 
\be
\langle u \rangle = \frac{ic}{\omega} e^{-i\omega t} \left[ 1 - \frac{\tan kd}{kd} \right].
\ee
The adiabatic, small $\omega$ limit leads to $ \langle u \rangle\sim \frac{c d^2}{3\nu}e^{-i\omega t} $
which is of Poiseuille type.
The opposite limit of fast frequencies leads to a plug-type velocity profile 
$ \langle u \rangle\sim \frac{ic }{\omega}e^{-i\omega t} $ with a thin boundary layer adjacent to the channel walls. 

\section{\label{sec: u1ACDC} First order in $\mathcal{N}$ correction to the velocity in the photogalvanic effect}

The first order dimensionless correction $u_1$ satisfies
\be \label{ut1}
\mathcal{R}\partial_\tau u_1 = \partial_y^2 u_1 + \frac{1}{2} \partial_y^2 u_0^2,
\ee
with no-slip boundary conditions at $y = \pm 1$. Because of the quadratic term in \rr{ut1}, the first order correction consists of a zero (DC) mode and an AC mode with twice the frequency $\omega$
\be
u_1(y, t) = u_1^{DC}(y) +  \frac{1}{2}u_1^{AC}(y)e^{-2i\omega t} +c.c.
\ee

Denoting the dimensionless amplitude in \rr{uLandau0} by $A(y)$
\be
A(y)=\frac{1}{2(kd)^2}  \left[\frac{\cos k y}{\cos kd}-1 \right]  
\ee
and by $A^*$ its complex conjugate, we obtain
\be \label{u02}
u_0^2 = - A^2 e^{-2i\omega t} + 2 |A|^2 - (A^*)^2 e^{2i\omega t}.
\ee
We consider below the DC term. The dimensionless Navier-Stokes become
\be
\partial_y^2\left[u_1^{DC} +  |A(y)|^2\right]=0
\ee
whose solution is
\be \label{u1dc}
u_1^{DC} = - |A(y)|^2. 
\ee

We also need the first order velocity that comes from the $2\omega$ harmonics in \rr{u02}. For the velocity $u_1^{AC} e^{-2i\tau}$, the dimensionless Navier-Stokes equations become 

\be
-2 i \mathcal{R} u_1^{AC} = \partial_y^2 u_1^{AC} - \partial_y^2 A^2,
\ee
or, since $\mathcal{R} = -i (kd)^2$, 
\be
 \partial_y^2 u_1^{AC} + 2(kd)^2  u_1^{AC} = \partial_y^2 A^2.
\ee
The general solution in dimensional variables is 
\be \label{u1AC}
u_1^{AC}  = u_p(y) - u_p(d) \frac{\cos \sqrt{2}k y}{\cos \sqrt{2}k d} 
\ee
where $u_p$ is the particular solution
\be
u_p(y) = U_0\frac{2\cos(ky)\cos(kd) + \cos(2ky)}{4(kd)^2 \cos^2 kd}. 
\ee
Collecting all the above into the velocity field \rr{uyt}, integrating over the width of the Hall-bar and taking the adiabatic limit $\omega\rightarrow 0$, it is easy to show that the net current reduces to \rr{IfNser}, which is the leading order expansion in $\mathcal{N}$ of the steady-state (zero vibrational number) current \rr{Iss}. 


\section{\label{sec: appendixdis} Effect of disorder-induced momentum relaxation}
In this Appendix we obtain the DC current discussed in section \ref{sec: disorder}, when weak disorder is present. 
The momentum equation \rr{ut0} acquires a disorder-induced friction force that is proportional to the velocity field of the electron liquid and takes the form \rr{NSdis}
\be \label{NSdis2}
\mathcal{R} \partial_\tau u = \cos\tau + \partial_y \left[ (1+ \mathcal{N} u)\partial_y u                 \right] - (k_{dis} d)^2 u.
\ee
We thus solve \rr{NSdis2} perturbatively with respect to $\mathcal{N}$. The velocity field retains the form \rr{uyt} and we have to determine the form of $u_0$ and $u_1^{DC}$ from which the DC current $I^{DC}_{dis}$ follows.

At $\mathcal{N} \equiv 0$ one obtains the 
velocity profile \rr{uLandau0} 
\be \label{uLandau0dis}
u_0 = \Re \left\{ \frac{-en\mathcal{E}}{\eta K^2} e^{-i\omega t} \left[ 1 - \frac{\cos K y}{\cos Kd} \right]   \right\}
\ee
where $k$ is replaced by $K = \sqrt{\frac{i\omega}{\nu} -  k^2_{dis} }$ (cf. Eq. \rr{K}). 
Denote the dimensionless amplitude in \rr{uLandau0dis} by $A(y)$
\be
A(y)=\frac{1}{2(Kd)^2}  \left[\frac{\cos K y}{\cos Kd} -1\right]  .
\ee
We consider below the DC term. The dimensionless Navier-Stokes become
\be
\partial_y^2\left[u_1^{DC} +  |A(y)|^2\right] - (k_{dis}d)^2 u_1^{DC} = 0. 
\ee
This is a second order equation with an unconventional nonhomogeneous term. To solve is easier to define the new function $v(y) = u_1^{DC} +  |A(y)|^2$ which satisfies 
\be \label{eqv}
\partial_y^2v  - (k_{dis}d)^2 v = -(k_{dis}d)^2  |A(y)|^2 
\ee
which we solve with variation of parameters. 
The dimensional DC velocity $u_1^{DC}$, satisfying no-slip boundary conditions at the Hall-bar lateral walls located at $y=\pm d$, becomes 
\be \label{u1dcdis}
u_1^{DC} = - |A(y)|^2 + v_p(y) - v_p(d) \frac{ \cosh k_{dis}y}{ \cosh k_{dis} d}.
\ee
where 
\be
v_p(y)  = k_{dis}^2 U_0\int  dy'  |A(y')|^2 G(y,y')
\ee
is the particular solution of \rr{eqv}
which is even about the origin $y=0$, 
and $G$ is the Green's function
\be
G(y,y') = \frac{ 
\left|\begin{array}{cc}
\cosh k_{dis} y & \sinh k_{dis} y\\
\cosh k_{dis} y' & \sinh k_{dis} y'
\end{array} \right|
}{W}, 
\ee
and the Wronskian is $W = k_{dis}$.

The net disordered DC current 
\be \label{IDCdis}
I^{DC}_{dis} =\!\!\! \int_{-d}^d\!\! en\mathcal{N} u_1^{DC}(y) dy
\ee
becomes
\begin{align}
 I^{DC}_{dis}  = &\frac{3I_0\mathcal{N}}{2qp^2d^5 |K|^6(4p^4 + 3q^4)} \left[ p^2\left(|\mathcal{T}|^2q^2 - 2|K|^2\right) \right.     \nonumber \\
 & \left. + 2qp^2 \Re (K\mathcal{T}^*) + 3q^3 \Im (K \mathcal{T}^*) \right], 
 \label{IDCdis2}
\end{align}
where
\be
p^2 = \frac{\omega}{\nu}, \quad q = k_{dis}, 
\quad \mathcal{T} = \tan Kh. 
\ee
We discuss the asymptotics of current \rr{IDCdis2} in section \ref{sec: disorder}. 


\section{\label{sec: v2} Nonreciprocal inertia}
\noindent 

In the absence of Galilean invariance, inertial properties of the liquid are characterized by a tensor $\rho_{ij}$, which relates the momentum density  to the hydrodynamic velocity, 
\begin{align}
    p_i = & \, \rho_{ij} v_j.
\end{align}
In nonreciprocal fluids, symmetry allows the inertia to acquire a linear in velocity correction
\begin{equation}\label{generalnrinertia}
    \rho_{ij} = \rho^{(0)}_{ij} + \rho^{(1)}_{ijk}v_k.
\end{equation}
Within the Fermi liquid picture, the nonreciprocal component of inertia can be expressed in terms of nonreciprocity of the electron spectrum, $\epsilon({\bf p}) - \epsilon(-{\bf p}) \neq 0$, by evaluating the momentum density of the liquid 
\begin{equation}
\label{eq:momentum_density_def}
    \bm{p} =  2\int \frac{d^2p}{(2\pi \hbar)^2} \  \mathbf{p} f(\mathbf{p}),
\end{equation}
for the equilibrium Fermi distribution with hydrodynamic velocity $\bm{v}$, 
$f(\mathbf{p}) = \left( e^{[\epsilon(\mathbf{p}) - \mathbf{p}\cdot \bm{v}]/T}+1\right)^{-1}$. Here the energy $\epsilon(\mathbf{p})$ is measured relative to the Fermi energy. 

As an illustration, we consider valley-polarized graphene~\cite{Zhou2021} with a small trigonal warping of the electron spectrum (which we measure from the Fermi energy),
\begin{equation}
\epsilon = v_F(p-p_F)+\Delta\cos(3\phi).\label{eq:trigonalwarpingspectrum}
\end{equation}
Here  $\phi$ is the angle between the electron momentum and $x$ axis, $\Delta \ll \epsilon_F$ is the energy of trigonal warping, and $p_F$ is the Fermi momentum at $\Delta=0$. For small temperatures (as compared to the Fermi energy) the evaluation of the integral in Eq. \eqref{eq:momentum_density_def} yields, 
\begin{subequations}\label{eq:pxy-corrections}
\begin{align}
    p_x &= \rho v_x + \rho_1\left(v_y^2-v_x^2\right),\\
    p_y &= \rho v_y + \rho_1\left(2v_xv_y\right),
\end{align}
\end{subequations}
where $\rho_1 \propto  \rho(\Delta/v_F\epsilon_F)$.

\begin{figure}[t!]
\includegraphics[width=\linewidth]{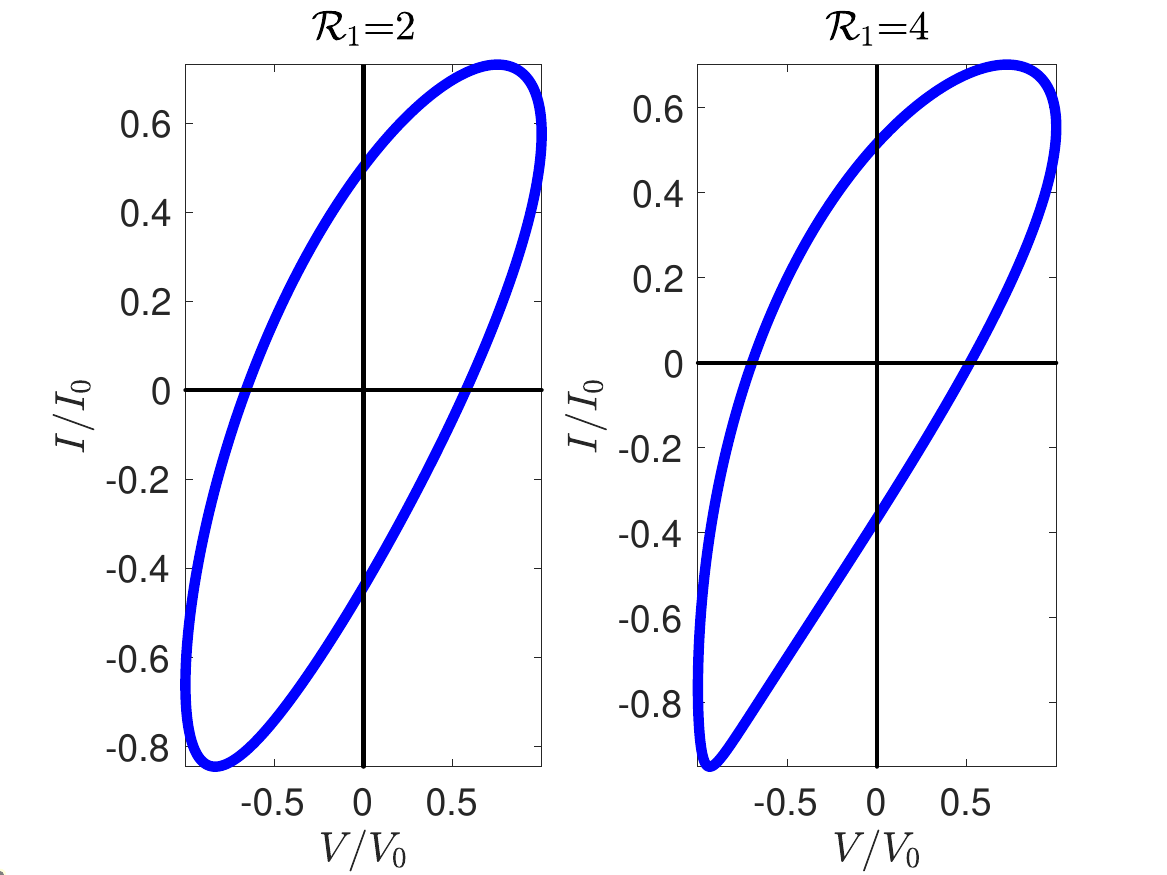}
\caption{Current-voltage hysteresis loops evaluated at $\mathcal{R}=2$ and $\mathcal{N}=0.1$. Increasing $\mathcal{R}_1$ in Eq. \rr{utR1}, distorts the loop of the left panel of Fig. \ref{memory1} (where $\mathcal{R}_1\equiv 0$).
\label{memory4}}
\end{figure}

Next, we show that the nonreciprocal component of the inertia of the electron liquid does not affect the DC component of the current described in the main body of this paper, and only affects the oscillatory component.
In the Hall bar geometry where only $x$-component of the flow is present,
the nonreciprocal correction to the momentum density given by Eqs. \eqref{eq:pxy-corrections}  becomes (per convention of the main text $v_x\to u$):
\be 
\label{sigmaR1}
\delta p \sim  \rho u \left(\frac{u}{v_F}\right)\left(\frac{\Delta}{\epsilon_F}\right).
\ee

It is convenient to define the velocity scale
$U_1 = v_F (\epsilon_F/\Delta)$ and the corresponding dimensionless parameter $\mathcal{R}_1 = \mathcal{R} (U_0/U_1)$,  
where $U_0$ is the velocity scale \rr{U0N} and $\mathcal{R}$ is the vibrational number \rr{NRev}. With \rr{sigmaR1}, the momentum equation replacing \rr{ut0}, becomes
\begin{equation}\label{utR1}
(\mathcal{R} + \mathcal{R}_1 u) \partial_\tau u = \cos\tau + \partial_y \left[ (1+ \mathcal{N} u)\partial_y u \right]. 
\end{equation}
The relative importance of the two
nonreciprocal terms in \rr{utR1} is determined by the ratio
$\mathcal{R}_1/\mathcal{N}=\mathcal{R}$ .

At $\mathcal{N}=0$, the second term on the left-hand side of Eq. \rr{utR1} does not produce a DC current.  This can be seen by writing the hydrodynamic velocity in the form $u=Z(y) + F(\tau,y)$  where $Z$ is the DC part (zero mode) and $F$ has a zero time-average and therefore, can be written as a time-derivative of a $2\pi$-periodic function of $\tau$, $F(\tau,y) =\partial_\tau Y(\tau,y) $. Averaging Eq.~\eqref{utR1} over the period of oscillation at $\mathcal{N}=0$ necessarily leads to the condition that $Z\equiv 0$.

The absence of the zero mode can also be established by expanding $u$ in a perturbation series with respect to a small parameter $\mathcal{R}_1$. The zeroth order solution is again \rr{uLandau0} but the first order corrections do not include a zero mode since the time derivative of $u_0^2$ in \rr{u02} contains only $2\omega$ harmonics. 

In Fig. \ref{memory4} we display current-voltage hysteresis loops obtained by numerically solving Eq. \rr{utR1}. Thus, the loops become skewed, showing the dependence of the oscillatory velocity field on $\mathcal{R}_1$.

\bibliography{biblio}

\end{document}